# Nontrivial Pure Zero-Scattering Regime Delivered by a Hybrid Anapole State

**Adrià Canós Valero[1], Egor A. Gurvitz[1], Fedor A. Benimetskiy[1], Dmitry A. Pidgayko[1], Anton Samusev[1], Andrey B. Evlyukhin[1], Mohsen Rahmani[2], Khosro Zangeneh Kamali[2], Alexander A. Pavlov[3], Andrey E. Miroshnichenko[4] and Alexander S. Shalin[1]**

[1]ITMO University, Kronverksky prospect 49, 197101, St. Petersburg, Russia

[2]Nonlinear Physics Centre, Research School of Physics, The Australian National University, Canberra, ACT 2601 Australia

[3]Institute of Nanotechnology of Microelectronics of the Russian Academy of Sciences (INME RAS), Moscow, Nagatinskaya street, house 16A, building 11

[4]School of Engineering and Information Technology, UNSW Canberra, ACT, 2600, Australia

**Abstract.** The ability to manipulate simultaneously electric and magnetic components of light at the nanoscale is paving the way to a plethora of new optical effects. Dielectric and semiconductor components enable novel types of sources and nanoantennae with exceptional electromagnetic signatures, flexible and tunable metasurface architectures, enhanced light harvesting structures, etc. Recently, the "anapole" states arising from the destructive interference of basic multipoles and their toroidal counterparts have been widely exploited to cancel radiation from an individual scattering channel of isolated nanoresonators, while displaying nontrivial near fields. As such, anapole states have been claimed to correspond to non-radiating sources. Nevertheless, these states are always found together with high order multipole moments featuring non-zero overall far-field. In this paper, we theoretically and experimentally demonstrate a fully non-scattering state governed by a novel 4-fold hybrid anapole with all the dominant multipoles of a disk nanoparticle being suppressed by their corresponding toroidal (retarded) terms. Such invisibility state, however, allows for non-trivial near-field maps enabled by the unique interplay of the resonant Mie-like and Fabry-Perot modes as demonstrated by the quasi-normal modal expansion. Moreover, the hybrid anapole state is shown to be "protected": the spectral position of the non-scattering point remains unperturbed in the presence of a substrate with significantly high refractive index. This, in its turn, enables the design of hybrid anapole based invisible metasurfaces disturbing neither phase nor amplitude of the incident wave but possessing strong near-field enhancement. We experimentally verify our novel effect by means of dark-field measurements of the scattering response of individual nanocylinders and obtain perfect agreement. The results are of high demand for efficient sensing and Raman scattering setups with enhanced signal-to-noise ratio, highly transmissive metasurfaces for phase manipulation, holograms, ultrafast computing, and a large span of linear and non-linear applications in dielectric nanophotonics.

## Introduction

Over the past few years, all-dielectric nanophotonics has become one of the cornerstones of modern research in nano-optics[1]. Unlike plasmonic structures, dielectric ones allow overcoming the fundamental limitation of Ohmic losses. Utilizing electric and magnetic Mie-like resonances of nanoparticles consisting of low-loss high-index semiconductor or dielectric materials, such as Si, $TiO_2$, Ge, $GaAs^{2,3}$, enables manipulating both the electric and magnetic components of light at the nanoscale. This emerging field has already led to a wide range of exciting applications, such as low-loss discrete dielectric waveguides[4,5] different types of dielectric nanoantennas (Yagi-Uda[6], super-directive antennas[7], etc.), directional sources[8], light-harvesting and antireflective coatings[9–12], all-dielectric metasurfaces with artificially tailored optical response[13–16], to mention just a few.

Therefore, the ability to properly describe and predict electromagnetic scattering is of prime importance to manipulate the behavior of light at the nanoscale. For this purpose, different types of electromagnetic multipole expansions were introduced[17–21]. Among them, the charge-current Cartesian decomposition is commonly and widely used for describing optical signatures of nano-objects of arbitrary shape[17–19]. One of the most intriguing possibilities delivered by this formalism is the ability to define the so-called toroidal moment family[22–25] allowing for additional flexibility in light governing. The electric toroidal dipole moment is the lowest order member of the toroidal family. Its formal meaning is associated with the poloidal currents flowing along the meridians of a torus and generating a magnetic field loop [23]. Higher-order toroidal moments, also known as mean square radii, feature more complex current distributions recently investigated in Refs.[26,27].

After the first experimental demonstration of artificial metamaterials presenting significant electric toroidal dipole contributions [25] toroidal moments began to attract significant attention. They are now widely exploited in nanophotonics and metamaterials, active photonics[28], ultrasensitive biosensing [22] and applications requiring strong near field localization[29,30]. The fields radiated by toroidal moments share the same angular momentum and far-field properties as their Cartesian electric or magnetic multipolar counterparts, allowing for the realization of two exciting effects: enhanced multipolar response [31] enabled by the constructive interference of the fields, and mutual cancellation of the far-field contributions via the destructive one, so-called "anapole" states. The electric dipole anapole was recently observed in a dielectric nanodisk[32].

In the aforementioned scenario, the far-field radiation from a given electric or magnetic multipole is suppressed. Consequently, anapole states, in principle, could drive low-scattering regimes[33,32]. This promising feature is of high demand for, e.g., cloaking and invisibility research[34–37] and a variety of nonlinear applications[38–41] enabled by strong near-fields. Very recently, they were also shown to generate energy 'spikes' in their transient response to an exciting pulse[42], a novel effect that could be of high relevance for the pioneering field of ultrafast dynamic resonant phenomena[43–45].

However, the great majority of the investigations on anapole-states are limited to the electric dipole term only[46–49]. Recently, the authors of Ref.[50], pointed out the exciting possibility to simultaneously cancel more than one multipole in spherical dielectric particles, which can be well understood in terms of Mie theory. However, in this case, "hybrid" anapole states are always hidden by the contributions of other multipole moments with non-negligible scattering[50], which is a strict limitation of the spherical geometry.

Fortunately, recent developments in the theory of multipole expansions[18,27] have opened the possibility to qualitatively and quantitatively investigate higher order electric and magnetic anapole states in scatterers with arbitrary shape, and allow their physical interpretation in terms of the interfering higher order electric and magnetic multipoles and their corresponding toroidal counterparts.

Here we demonstrate for the first time the existence of such exotic dark states, schematically illustrated in **Figure 1**. We theoretically predict a novel type of non-trivial invisibility (non-scattering regime), accompanied by an effective internal field concentration, governed by the spectral overlap of four electric and magnetic anapole states of different orders. This pure *Hybrid Anapole* state originates from the far-field destructive interference of *all* the leading electric and magnetic Cartesian multipoles of a finite cylindrical scatterer with their associated toroidal moments. Complementing the new multipolar approach, the near-field maps are interpreted in terms of the fundamental modes of an open cavity[51,52] (quasi-normal modes), thus providing a complete physical picture of the effect. Moreover, we experimentally observe these hybrid anapole states in the visible range (with excellent agreement with the theoretical predictions), in high-index silicon nanoparticles with their sizes comparable to the wavelength of the incident light.

The manuscript is organized as follows: we start by giving a short overview of the approach utilized for the calculation of the multipoles providing a quantitative description of high order anapole states in terms of the interference between basic and toroidal moments. Then we proceed to the particular situation of a four-fold anapole state realized in a high-index cylindrical nanoparticle. The physical origin and the characteristic spectral features of the effect are then discussed in terms of both the irreducible multipole and quasinormal mode expansions. Next, the influence of different dielectric substrates on the near- and far-fields is investigated. Moreover, we investigate a metasurface consisting of the hybrid anapole particles, showing its full invisibility in terms of phase and amplitude of the transmitted light. Finally, we provide a description of the experimental setup, details of the measurements and a direct comparison between the theoretical and experimental results.

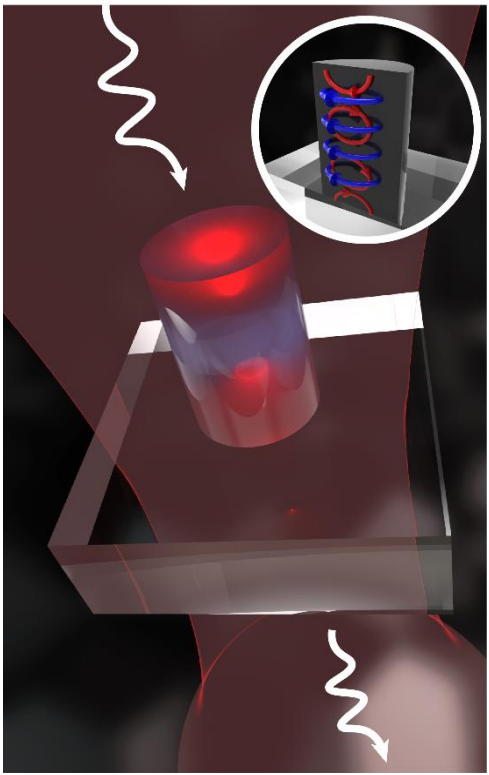

**Figure 1.** Artistic representation of the novel effect. A normally incident wave excites nontrivial modal contributions in a Si nanocylinder whose interference with the background field leads to a four-fold hybrid anapole state (see **Figure 2**a), yielding the nanoantenna virtually invisible in the far-field, with strongly localized near field. The same eigenmodes are responsible for enhanced electromagnetic energy stored inside the resonator. Inset depicts the current distributions of the two resonant eigenmodes arising due to standing waves between the top and bottom walls (red) and lateral walls (blue) in the vertical plane.

**The Cartesian multipole expansion and high-order anapole conditions**

The analysis of the optical response of a nanoparticle is usually carried out via the decomposition of the scattering cross section as a sum of multipoles, which represent independent scattering channels of the object. Here we utilize the irreducible Cartesian multipole expansion derived in Ref.[27] (for completeness, also given in the Supplementing Information), that explicitly takes into account higher order toroidal moments. The latter interpretation is our starting point towards the physical understanding of higher-order anapole states.

Within this approach, an electric or magnetic anapole of order *n* in a subwavelength scatterer is given by the condition:

$$P^{(e,m)}_{i_1...i_n} + i\frac{k_d}{v_d}T^{(e,m)}_{i_1...i_n} = 0 \qquad (1)$$

Here we have denoted *n*th order electric or magnetic moments with $P^{(e)}$ and $P^{(m)}$ and corresponding electric and magnetic toroidal moments with $T^{(e)}$ and $T^{(m)}$, respectively. The number of subscripts indicates the order of each Cartesian tensor, i.e., one subscript corresponds to dipole, two correspond to quadrupole, etc. $k_d, \varepsilon_d$ are the wavenumber and the dielectric permittivity of the host medium, and $v_d$ is the speed of light in the medium. A hybrid anapole state occurs when more than one multipole moment fulfils Eq.(1) at a given wavelength, resulting in a simultaneous suppression of scattering of two or more channels. However, as mentioned above, light, in general, can be radiated out through other non-zero multipole moments, destroying the overall effect. Thus, only the cancellation of all the leading multipoles can enable true low-scattering or "invisible" regimes.

For the sake of clarity, in the rest of the manuscript, we will rely on the widespread notation for low-order multipoles, i.e. $p$, $m$ for electric and magnetic dipoles, and $Q^{(e)}$, $Q^{(m)}$ for electric and magnetic quadrupoles.

**Near-zero scattering hybrid anapole states**

Under conventional plane wave illumination, hybrid anapole states of homogeneous spherical particles are hidden by the contributions of high order multipoles [50,53]. Note that this restriction naturally vanishes for nanoobjects with additional geometrical degrees of freedom, like finite cylinders or parallelepipeds [54]. Throughout this work, and particularly in the next section, we will unveil the fundamental reason behind this unusual behavior.

Let us consider a cylindrical silicon nanoparticle in air. Starting now, we will use amorphous silicon (aSi) in both theoretical and experimental studies (the dielectric function measured from the unstructured thin film can be found in the Supplementing Information). The illumination scheme is presented in the left inset of **Figure 2**a (normally incident *x*-polarized plane wave propagating along $-z$ direction).

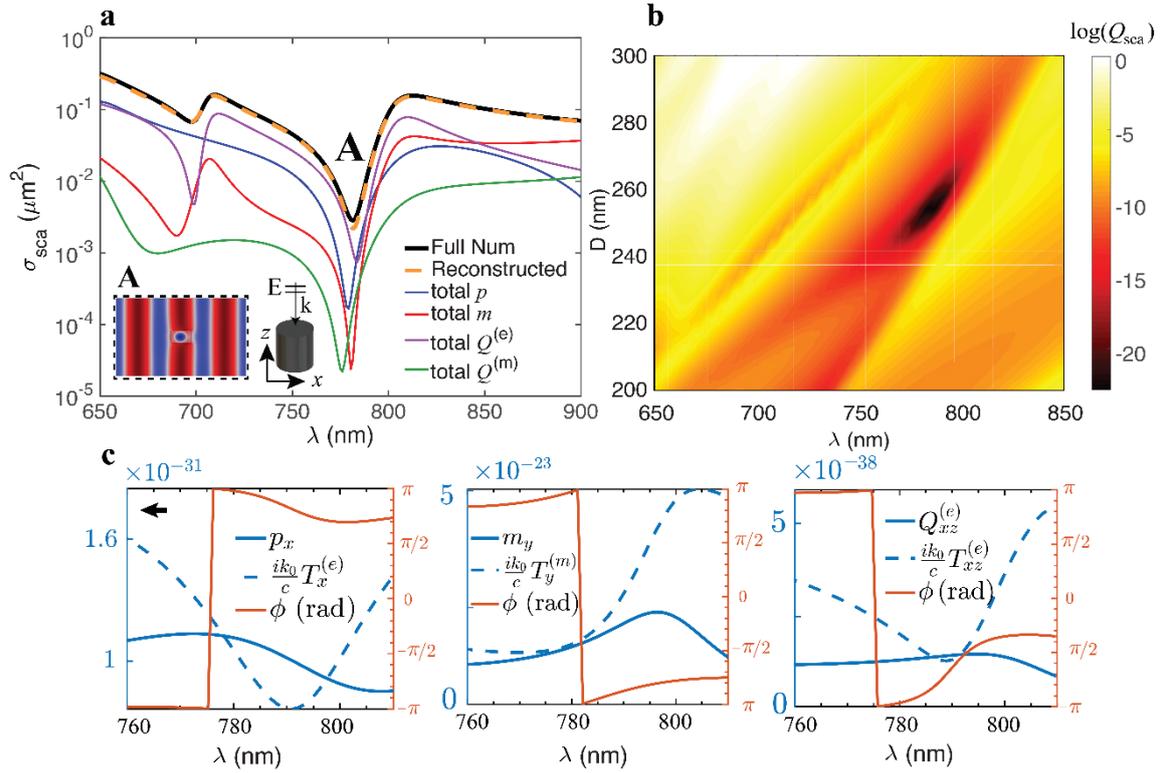

**Figure 2.** (a) Multipole reconstruction of the numerically obtained scattering cross section for the cylindrical amorphous silicon nanoparticle (see Supplementing Information for the measured refractive index dispersion). In the legend caption, "total" implies that both basic and toroidal contributions of a given multipole are plotted. Left inset of (a) corresponds to the x-component of the electric field near the nanoparticle, right inset schematically depicts the cylindrical particle and the illumination setup. The geometrical parameters of the cylinder are height H=367 nm, diameter D=252 nm. Point A ($\lambda = 782$ nm) corresponds to the hybrid anapole state. (b) Dimensionless scattering efficiency $Q_{sca} = 4\sigma_{sca}/\pi D^2$ map as a function of the diameter. Each panel in (c): Amplitudes and phase differences between the multipoles and their toroidal counterparts; from left to right, respectively, the basic electric and electric toroidal dipoles, the basic magnetic and magnetic toroidal dipoles, and the basic electric and electric toroidal quadrupoles. Amplitudes correspond to the left ordinate-axis, and phase differences are read from the right ordinate-axis.

The design methodology is based on the following: firstly, the excitation geometry restricts the excited electric multipole components to the plane of incidence (k-E plane), and the magnetic multipoles to the perpendicular k-H plane, (see Supplementing Information). The incident field induces, respectively, the $p_x$ and $Q_{xz}^{(e)} = Q_{zx}^{(e)}$ components of the electric dipole and quadrupole moments, and the $m_y$ and $Q_{yz}^{(m)} = Q_{zy}^{(m)}$ components of the magnetic dipole and magnetic quadrupole, respectively. Toroidal multipoles follow the same rule as their basic counterparts, and a similar scenario takes place for higher-order multipoles (see Supplementing information for a rigorous derivation).

Secondly, we note that the spectral positions of the full (basic and toroidal parts) electric dipole and magnetic quadrupole anapoles are mainly dependent on the cylinder's radius, while the wavelengths of the full magnetic dipole and electric quadrupole anapoles change both as functions of the cylinder height and radius. Figure S3 in the Supplementing Information illustrates the spectral behavior of the multipolar anapoles with variations in the height and diameter of the cylinder in detail. Thus, carefully tailoring these two geometrical degrees of freedom makes possible to place the anapoles of all the leading terms ultimately close to each other (**Figure 2**b), providing a strong scattering minimum (**Figure 2**a, point A).

The total scattering cross section and its multipole decomposition after the numerical optimization are shown in **Figure 2**a. Perfect agreement between the sum of the multipole contributions given by Eq.(1) in the Supplementing Information and the result of the full-wave simulations in Comsol Multiphysics is demonstrated, proving that only the first four multipoles are sufficient to fully describe the optical response of the nanocylinder in the visible range. Therefore, the low-scattering regime delivered by this state provides almost perfect particle invisibility in terms of far-fields (see the right inset on **Figure 2**a).

The different panels in **Figure 2**c show the amplitudes and phase differences of the three most relevant multipoles with their toroidal moments. The results further confirm that the generalized anapole condition in Eq.(1) is well fulfilled for each pair (the amplitudes are equal, and they are $\pi$ rad out of phase) at the hybrid anapole wavelength $\lambda = 780$ nm.

Noteworthy, the near-zero values of the full scattering coefficients do not imply the induced polarization currents in the particle to be also close to zero. This is in agreement with the usual anapole behavior [55], and, due to the suppression of several multipoles simultaneously, the hybrid anapole also displays well confined internal fields. **Figure 3**b demonstrates the average electromagnetic energy density inside the cylinder at the hybrid anapole wavelength to exceed 9 times the value of free space almost without leakage of the field outside the particle volume. This, in its turn, further enhances invisibility (see Figure S1a in the Supplementing Information) and reduces the interaction with the surrounding (see the following sections).

**Quasi-normal mode analysis of the hybrid anapole state**

While Cartesian multipoles are suitable for the description of far-fields, in this section we employ the natural quasi-normal mode (QNM) expansion[52] of near-fields and internal currents, which in the following will allow us to further unveil the physics behind the hybrid anapole state. QNMs provide a suitable basis for the induced polarization currents:

$$\mathbf{J}(\omega,\mathbf{r}) = \sum_s \alpha_s(\omega)\tilde{\mathbf{J}}_s(\omega,\mathbf{r}) - i\omega\varepsilon_0(\varepsilon_p - \varepsilon_m)\mathbf{E}_{inc}(\omega,\mathbf{r}). \qquad (2)$$

Here $\tilde{\mathbf{J}}_s(\mathbf{r}) = -i\omega\varepsilon_0(\varepsilon_p - \varepsilon_m)\tilde{\mathbf{E}}_s(\mathbf{r})$ and $\alpha_s(\omega)$ are, respectively, the induced modal scattering current distribution as a function of the internal mode field, and the excitation coefficient of the *sth* mode describing its contribution to the total current at a given frequency.

We use a modified version of the freeware MAN developed by the authors of Ref.[51]. More details on the approach can be found in the Supplementing Information. For simplicity, we consider a dispersionless, lossless nanocylinder with a constant refractive index $n \approx 3.87$ (corresponding to aSi at 780 nm), so that the excitation coefficients depend solely on the fields of an individual QNM[51]. Losses and refractive index dispersion of the original design are negligible in the considered spectral range (see Supplementing Information), and therefore this approximation does not remarkably change the scattering cross section and average electromagnetic energy density.

The results of the QNM expansion are displayed in the different panels of **Figure 3**. The correctness of our calculations in the studied spectral range, particularly near the scattering dip, is well validated in **Figures 3**a-b by comparing the sum of the individual QNM contributions with the numerically obtained total scattering cross section (a) and average electromagnetic energy density inside the cylinder (b). From hereon we shall label the QNMs with the standardized notation for the modes of isolated cylindrical cavities[56], i.e. $(TE,TM)_{uv\ell}$, where the sub-indices denote the number of standing wave maxima in the azimuthal ($u$), radial ($v$) and axial ($\ell$) directions. *TE* and *TM* indicate the predominant nature of the internal field distribution. Specifically, *TM* (transverse magnetic) modes have $H_z \approx 0$, while *TE* (transverse electric) have $E_z \approx 0$.

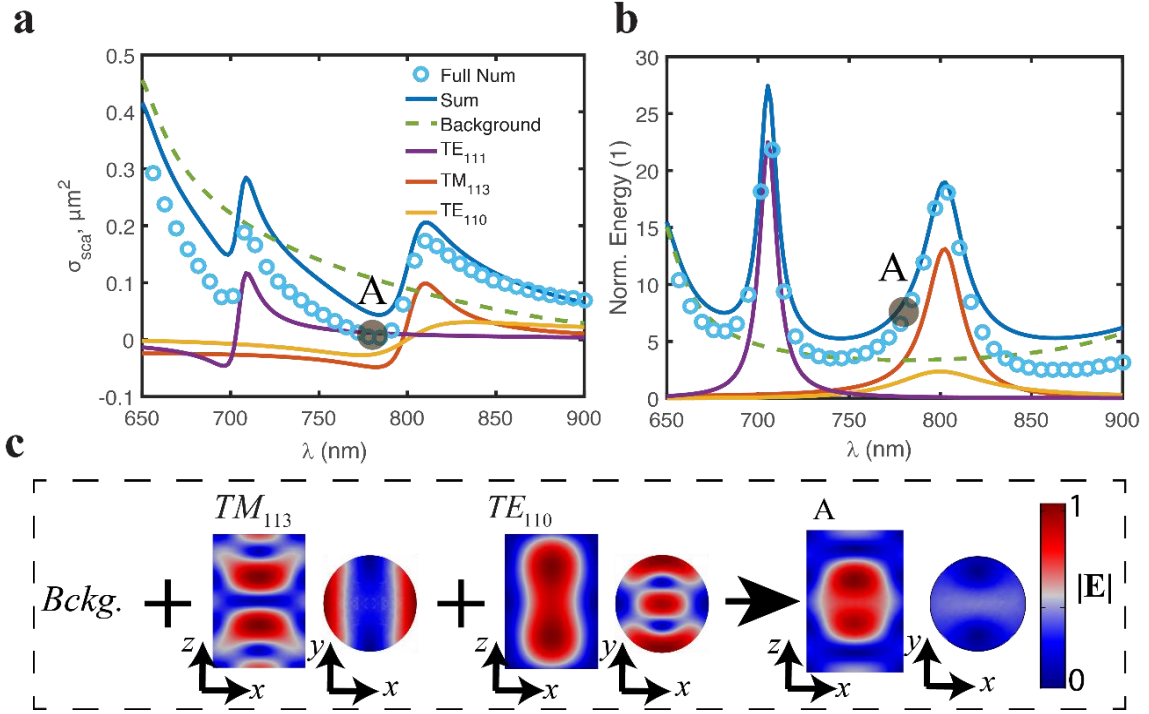

**Figure 3.** (a) Alternative scattering cross section decomposition by means of the QNM expansion method. The full-wave simulation is nearly the same (without losses) as in **Figure 2**, but in a linear scale. Colored lines are the individual contributions of the physical QNMs. The contributions of modes having their resonances outside the considered spectral range are added up in the green dashed line. Resonant modes in the considered spectral range are associated with the $TE_{111}$, $TE_{110}$ and $TM_{113}$ modes of the isolated cylinder. The blue line corresponds to the reconstructed scattering cross section, confirming that all the resonant spectral features can be successfully recovered via this method, and demonstrating good agreement near the hybrid anapole, point A. (b) Spectra of the volume-averaged electromagnetic field energy inside the cylinder, and individual contributions of the excited modes. Colors and legends are the same as in (a). The electromagnetic energy density has been normalized with respect to the incident electromagnetic energy density in vacuum $w_{EM} = \varepsilon_0 E_0^2$. Excellent agreement is obtained with the full-wave simulations. (c) Normalized internal electric field distributions of the two most relevant modal contributions near point A, from left to right, associated with Fabry-Perot ($TM_{113}$) and Mie-like ($TE_{110}$) standing wave patterns ($TE_{111}$ is very weak near anapole point), respectively. Their addition via Eq. (2), together with the background modes (Bckg.), leads to the reconstruction of the internal fields of the hybrid anapole, also displayed on the right-hand side of (c). All the electric fields have been normalized with their respective maxima, to enhance their visualization.

The spectral behavior of each resonant QNM can be well described by means of the well-known Fano formula with two critical points, representing maxima and minima in scattering [57], as demonstrated in section 5 of the Supplementing Information. The other nearby QNMs constitute the background scattering contribution of the particle.

In **Figure 3**a we note that a total of three QNMs strongly resonate in the visible range. The correct reconstruction of the scattering cross section requires taking into account background modes, despite their resonances being outside the considered spectral range (green dashed line). Nevertheless, at point A, only the $TM_{113}$ and $TE_{110}$ modes present a resonant 'Fano-like' response.

Now the invisibility effect can be easily grasped as a consequence of modal interference: a clear sign that this is indeed the case are the resonant negative contributions to scattering presented by both the $TE_{110}$ and $TM_{113}$ modes. It implies that, when the incident field impinges in the resonator, energy exchange takes place between the two and the background QNM fields[58]. This owes to the fact that the QNMs do not obey the usual conjugate inner product relation of orthogonal modes in Hermitian systems[59]. Here it is important to emphasize the unusual feature of the hybrid anapole: the two resonant QNMs dominating the spectra are *simultaneously negatively suppressed* by interference with the background. For comparison, the QNM decomposition of a conventional dipole anapole disk is given in the Supplementing Information.

A completely different picture arises within the resonator. **Figure 3**b presents the modal decomposition of the internal energy stored in the cylinder in the vicinity of point A. This is one of the main results of the section, since, contrarily to the multipole expansion, the QNM decomposition allows us to clearly distinguish the contributions of the different modes to the internal fields. Firstly, we note that similarly to an electric dipole anapole particle[32], the electromagnetic energy is significantly enhanced (around 9 times) with respect to the incident plane wave. Secondly, it is clearly seen that the stored energy at the hybrid anapole is mainly driven by the $TM_{113}$ mode due to its higher quality factor and the proximity of its resonant wavelength to the hybrid anapole wavelength, and in a minor measure by the $TE_{110}$ and the sum of the background contributions. Energy exchange between the internal fields of the QNMs is strongly minimized, as reflected in the fact that no negative contributions to the internal energy can be appreciated. Overall, the QNM analysis given in **Figure 3** demonstrates that both the invisibility effect (outside the cylinder) and the internal energy enhancement at the hybrid anapole state are mediated by the simultaneous resonant response of the $TM_{113}$ and the $TE_{110}$ modes. The background modes, on the other hand, while they do not apparently define the spectral features of the figures of merit significantly, play also an important role since their interference with the resonant ones gives rise to the invisibility effect. This interpretation is consistent with early investigations regarding the formation of Fano lineshapes in the scattering cross section of spherical resonators[53].

The electric field distributions of the $TE_{110}$ and $TM_{113}$ modes are shown in **Figure 3**c. Following Refs.[60,61] we can classify the first as a 'Mie' type mode, similar to the ones supported by an infinite cylinder, while the second is of the 'Fabry-Perot' (FP) type [60], arising due to the formation of a standing wave pattern between the top and bottom walls of the resonator, i.e. having non-zero axial wavenumber ($\ell \geq 1$). Their distinct origin unveils the reason why it is possible to obtain a hybrid anapole state in this particular geometry, contrarily to spherical scatterers. As shown in the Supplementing Information, the real parts of the eigenwavelengths of the modes in the cylinder can be estimated as[62]

$$\lambda_{uv\ell} \approx \frac{\pi D}{n_p \sqrt{\left(\frac{\ell \pi}{2} \frac{D}{H}\right)^2 + z_{uv}^2}}, \tag{3}$$

where $z_{uv}$ is the $v$th root of the $u$th Bessel function of the first kind for *TE* modes, or its first derivative for *TM* modes. For FP modes, $\ell \neq 0$ and the denominator in Eq. (3) displays a strong dependence on the aspect ratio $D/H$ of the cylinder. In contrast, since $\ell = 0$ for Mie modes, their eigenwavelengths only change with $D$. Thus, the eigenwavelengths and the multipolar content of these two mode types are independently tunable from each other, resulting in a flexible control over the optical response of the resonator and enabling the simultaneous scattering suppression observed in **Figure 2**a, and **Figure 3**a - the hybrid anapole state.

A straightforward comparison between the QNM and multipolar methods allows determining the multipoles radiated by a given QNM (see section 10, Figure S4, in the Supplementing Information). Specifically, Figure S3 leads us to the conclusion that the $TE_{110}$ mode radiates primarily as $p$, with a minor $Q^{(m)}$, while the $TM_{113}$ mode scatters as a combination of $m$ and $Q^{(e)}$ (in both cases, referring to both their basic and toroidal counterparts). When scattering from a mode is resonantly suppressed, radiation from the multipoles associated to it is also strongly minimized, and results in the different multipole anapole states observed in the decomposed spectra of **Figure 2**a. Thus, the close proximity of the destructive interference points of the $TE_{110}$ and $TM_{113}$ modes at point A leads to an overlap of the anapole states of the four dominant multipoles. In this fashion, using the QNM expansion approach, we have shown an alternative and general physical explanation of dark scattering states and qualitatively illustrated its link to the multipolar response of the particle.

As a final remark, we point out that the quality factors of the $TE_{110}$ and $TM_{113}$ modes in the structure are, respectively, 12 and 33. A direct comparison with the $TE_{110}$ supporting a conventional anapole disk (see Supplementing Information) shows that the quality factor of the $TM_{113}$ mode at the hybrid anapole is more than four times larger. Therefore, we anticipate much better performance of the hybrid anapole for second and third harmonic generation processes with respect to conventional anapole disks, since nonlinear scattering cross sections scale linearly with the quality factors of the modes involved[63].

Summarizing the results of this section, we have investigated the fundamental origin behind the hybrid anapole state and found that this novel effect arises from the modal interference of scattering by resonant FP-type and Mie-like modes excited in the resonator with the background off-resonant modes. Contrarily to a vast majority of the previous works on the conventional electric dipole anapole state, the modal decomposition has allowed us to study the separate contributions to the internal field, highlighting the connection between the irreducible Cartesian multipoles and the QNM expansion. We utilized the unique advantages of both approaches, delivering a unified physical picture of the effect in the near- and far-field regions. Importantly, the considered analysis and explanations developed here are general and can be utilized to understand the mechanisms behind anapole states of arbitrary order in resonators of arbitrary shape, once a sufficient set of the QNMs is known.

**Evolution of the hybrid anapole state in the presence of a substrate**

Realistic experimental conditions and possible applications require nanoparticles to be deposited on a substrate. Therefore, we should address and explain its influence on the hybrid anapole state.

In a general situation, the incident and scattered fields (including near and far-field contributions) are partially reflected back from the substrate inducing 'effective' multipoles, different from the case of a homogeneous environment [64]. The latter implies that any 'anomalous' scattering feature displayed by the nanoparticle, such as e.g., Kerker-like effects [65], relying on a delicate balance between the phases and amplitudes of the leading multipoles, will be perturbed. More formally, the substrate breaks inversion symmetry of the surrounding environment and introduces different possible multipole coupling channels in the effective polarizability tensor describing the particle response to an external electromagnetic field [64].

In comparison with an arbitrary spectral point, the hybrid anapole state is remarkably robust in the presence of a substrate. This particular behavior is illustrated in **Figure 4**a, where we have plotted the calculated scattering cross sections of the cylinder in free space and deposited over glass ($n_s = 1.5$), hypothetical substrates with $n_s = 2, 3$ and over amorphous silicon ($n_s = 3.87$). It can be appreciated that the amplitude and spectral position of the scattering dip are almost not displaced from the free-space values while the refractive index contrast at the bottom of the particle remains non-zero. However, we observe drastic changes in the shape of the Fano profile and anapole state efficiency, once the contrast is absent (silicon particle over silicon substrate). This effect suggests an underlying mechanism by which the hybrid anapole is "protected" against modifications of the substrate refractive index, that will be unveiled in the following.

Intuitively, the substrate invariance can be understood by considering the distinct nature of the two resonant modes involved in the formation of the hybrid anapole (**Figure 4**c). On the one hand, increasing the refractive index of the substrate leads to a decrease in the lower wall reflectivity, which is crucial for the standing Fabry-Perot mode $TM_{113}$ inside the resonator. The simulations show that an increase of $n_s$ results in a higher energy leak towards the substrate and a substantial decrease in the mode quality factor (see **Figure 4**c). Consequently, its resonance flattens and disappears when the lower boundary becomes transparent, as shown in **Figure 4**a.

The standing wave pattern of the $TM_{113}$ shows a close resemblance with the modes sustained by a one-dimensional Fabry-Perot resonator with the same height $H$ as the nanocylinder, deposited over a dielectric substrate. The QNMs of this simplified model have the advantage of being analytically solvable[52], thus providing valuable physical insight easy to extrapolate to the problem at hand. As we derive in the Supplementing Information, the QNMs are formed by two interfering plane waves traveling in opposite directions inside the cavity, when the driving wavelength satisfies the condition

$$r_{21} r_{23} w^2 = 1, \quad (4)$$

where $w = \exp(ik_\ell n_r H)$ and $r_{21}$, $r_{23}$ are the Fresnel reflection coefficients from the cavity-air and the cavity-substrate interfaces, respectively. The quality factor of a QNM with axial index $\ell$ is calculated as

$$Q_\ell = -\frac{\ell \pi}{\ln(r_{23} r_{21})}. \quad (5)$$

Equation (5) serves very well to illustrate the influence of the substrate on the $TM_{113}$ mode. To do so, it is also important to notice that the numerator is proportional to the real part of the eigenwavelength, while the denominator is linked to its imaginary part. When the two reflection coefficients are unitary, the energy is completely stored inside the resonator and the quality factor is infinite. Similarly, a decrease in the reflection coefficient from the substrate leads to radiative losses and a decrease in the quality factor, effectively becoming zero when the substrate refractive index matches that of the particle (i.e. no refractive index contrast). Indeed, a lower quality factor leads to less appreciable spectral features, as observed in the simulations (**Figure 4**a). Another important conclusion that one can draw from the numerator in Eq. (5) is that the real part of the resonant wavelength of the QNM is independent of the refractive index at the walls. Thus, modifying the substrate refractive index does not shift the spectral position of the resonance (i.e. does not shift the hybrid anapole wavelength), but simply changes the amplitude and width of the Fano profile.

With proper normalization, and employing the notation of the inset in **Figure 4**a, the amplitudes of the incoming and outgoing plane waves inside the resonator are $\left|A_2^+\right| = \sqrt{\alpha r_{12} w}$ and $\left|A_2^-\right| = \sqrt{\alpha r_{23} w}$, with $\alpha = 1/(4H\varepsilon)$ (see Supplementing Information for details). As could be expected, plane waves reflected from the substrate are decreased in amplitude when the refractive index contrast becomes lower. Consequently, the field maxima of the $TM_{113}$ mode closer to the substrate are reduced in amplitude. While this prediction is observed in the simulations for relatively low contrasts, the behavior at very small contrasts is very different (see the case with $n_s = 3$ in **Figure 4**c). In the latter situation, the standing waves along z become negligible in comparison with the initially weaker standing waves in the x-y plane, and the QNM can only be well described numerically.

Contrarily, it is noteworthy that even comparably small contrast (3 – 3.87) leads to enough contribution of the $TE_{110}$ mode to still preserve the scattering dip (see **Figure 4**a). The standing wave pattern of this second mode develops in the lateral walls of the cylinder, and therefore depends much less on variations of the reflectivity of the lower wall, keeping an almost constant quality factor (see Supplementing information). Most of the QNM energy is then stored in the resonator even in the case of zero effective contrast with the substrate, as demonstrated in **Figure 4**c. This results in a larger contribution of the $TE_{110}$ mode to extinction at small contrasts in comparison with the $TM_{113}$ mode. As a consequence, the electric quadrupole and magnetic dipole contributions are no longer dominant, and the hybrid anapole is mainly driven by electric dipole radiation stemming from the $TE_{110}$ mode (see Supplementing Information). Thus, for large $n_s$ the hybrid anapole *degenerates into a conventional electric dipole anapole state*, but still retains its non-radiative nature, since the other multipoles become negligible.

To conserve the four-fold hybrid anapole, it suffices that the particle-substrate interface reflectivity is high enough to support the standing waves in the z direction of the $TM_{113}$ mode. The panels in **Figure 4**b show the total (left) and scattered fields (right) corresponding to the hybrid anapole state with and without a glass substrate. Both designs demonstrate nearly perfect scattering cancellation - the scattered amplitude does not exceed 2% - at the close vicinity of the nanoparticle and is even much less in the far-field zone. Since the detectable signature of an object is the wave-front distortion (in terms of phase and/or amplitude), our cylinder clearly displays the features characterizing an invisibility state while still providing strong internal fields due to the excited modes.

The stability of the hybrid anapole state against changes in the substrate refractive index is an important result, since it renders the effect as an ideal building block for the practical implementation and study of non-linear phenomena, as well as for the development of novel functional fully transparent metasurfaces, enhanced Raman scattering setups, and a variety of applications requiring high signal/noise ratios.

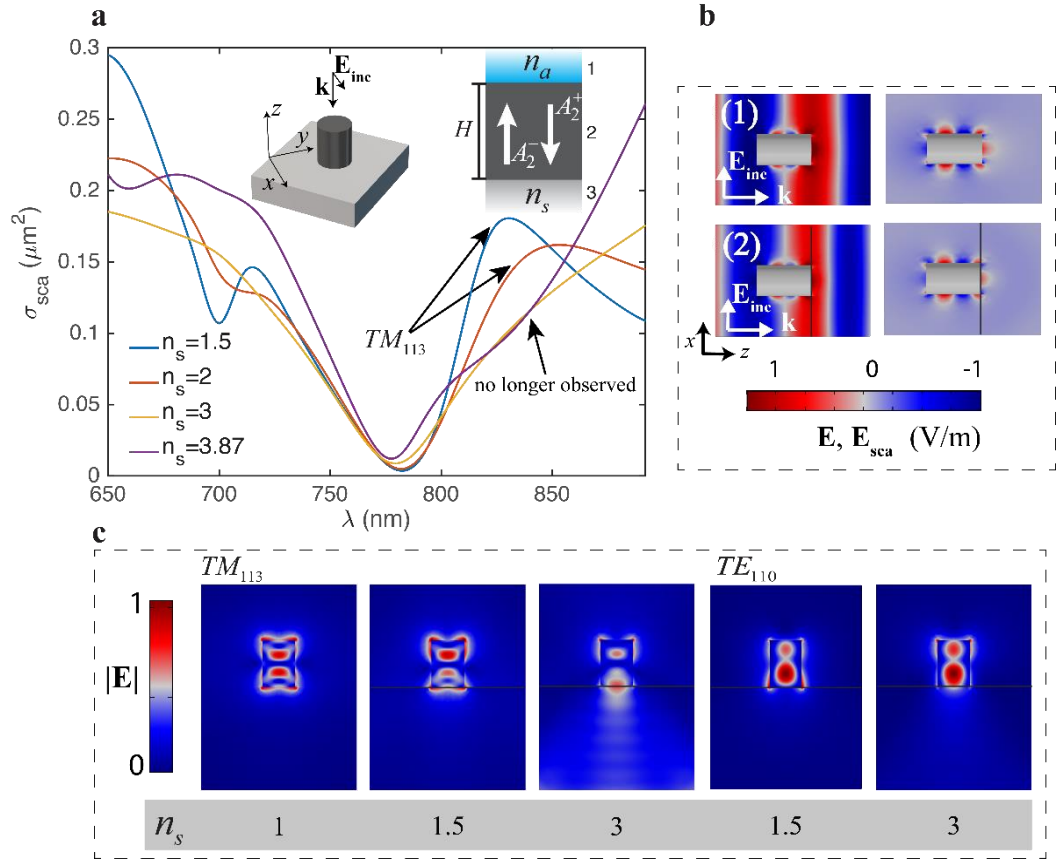

**Figure 4**. Evolution of the hybrid anapole state with different substrates (a) Comparison between the numerically obtained scattering cross sections for the nanocylinder with the size from **Figure 2**, deposited on substrates with increasing refractive index. The calculations are performed with the full experimental aSi refractive index given in the Supplementing Information. Left inset of (a): Schematic view of the particle placed on a substrate, under normally incident plane wave illumination. Right inset: One-dimensional Fabry Perot model of the $TM_{113}$ mode. (b) Electric field distribution for the indicated plane wave illumination: (1) the particle in free space, left panel: total field, right panel: scattered field $\mathbf{E}_{sca} = \mathbf{E} - \mathbf{E}_{inc}$ (2) Deposited on the glass substrate, left panel: total field, right panel: scattered field. The incident field amplitude is $E_0 = 1 \text{V/m}$. Colorbar is set equal for both total and scattered fields for better visualization. (c) Field distributions of the QNMs $TE_{110}$ and $TM_{113}$ when the cylinder is deposited over substrates with the different refractive index. As predicted by our qualitative theory, losses from the $TM_{113}$ mode increase when decreasing the refractive index contrast. Contrarily, the $TE_{110}$ mode remains confined in the scatterer.

Thus, we conclude that the changes in the Fano profile even in the limiting case of zero contrast can be well understood by investigating the dissimilar behavior of the two resonant QNMs involved. In this fashion, we have provided a simple and intuitive physical description of the mechanisms underlying the hybrid anapole behavior in the presence of dielectric substrates. The considerations in this section will be invaluable in the future to understand and design real-world applications based on the proposed effect.

**Experimental Observations**

To confirm our theoretical predictions on the novel hybrid anapole state, we have carried out direct scattering spectroscopy measurements for a set of single nanocylinders with the designed dimensions in the optical spectral range. The aSi structures were fabricated on a glass substrate using electron beam lithography (EBL) combined with plasma etching. First, a 367 nm thick aSi film was deposited on the glass substrate with plasma-enhanced chemical vapor deposition. Subsequently, circular masks out of Chromium (Cr) were fabricated on the films via EBL, followed by inductively coupled plasma etching to transfer them into the silicon film. Finally, residual Cr masks were removed via Cr etchant.

The scattering response of the individual nanoresonators was measured utilizing a home-built forward-scattering dark-field spectroscopy setup, schematically shown in **Figure 5**c. In the experiment, the sample is excited by a weakly focused white light beam incident at small angles (objective lens with $NA = 0.26$). The incident radiation passing through the sample without scattering is blocked in the intermediate back focal plane (BFP) in the collection channel. The light scattered by the nanoparticle in the forward direction was collected by a high-NA objective lens ($NA = 0.7$) and analyzed with a spectrometer.

The measured scattering spectra (solid lines, **Figure 5**a) exhibit a pronounced dip, shifting with the increase of the nanocylinder diameter D, in excellent agreement with the calculations (dashed lines, **Figure 5**a). As expected, since the Fano resonances governing the hybrid state are driven by different modes, they behave differently with changes in the geometry (see Supplementing Information). It results in modifications of the amplitude and the overall shape of the scattering efficiency dip with variations in the cylinder diameter. While an increase of the lateral size of the nanocylinder leads to an overall redshift of the multipole anapoles, they overlap at $D = 251$ nm, forming the hybrid anapole state. This leads to the most pronounced scattering efficiency dip, of almost two orders of magnitude, rendering the nanocylinder virtually invisible.

The observations undoubtedly prove the existence of hybrid anapole states in our configuration, which can be potentially exploited for a large span of applications in linear and nonlinear optics.

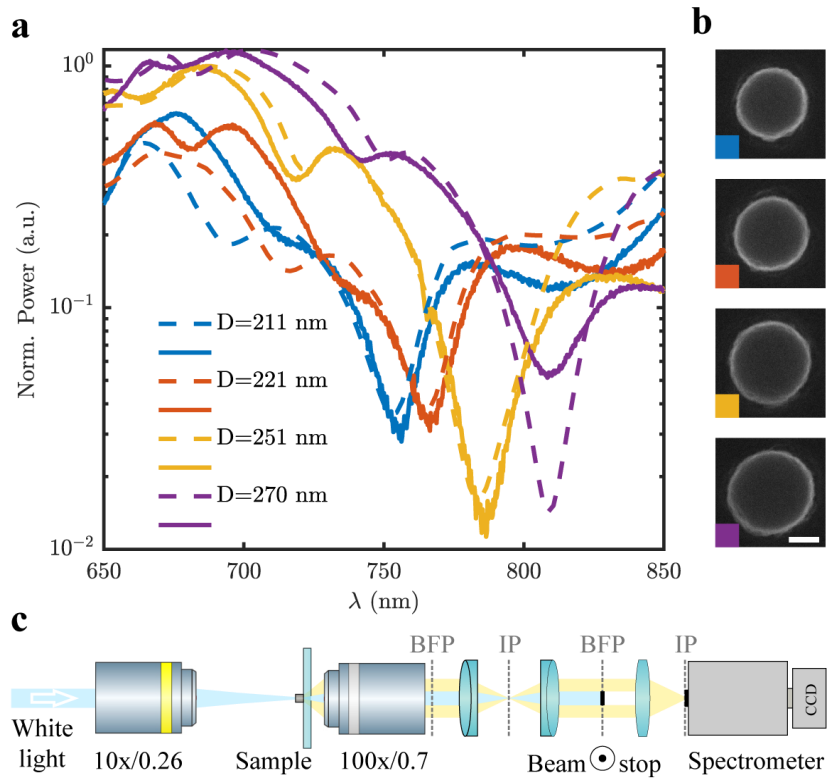

**Figure 5.** (a) Measured (solid lines) and simulated (dashed lines) scattering spectra of single isolated nanocylinders with different diameters D. (b) SEM micrographs of the corresponding nanocylinder samples. The colored edges in each micrograph are associated to the legend entries in (a). A faint bright tone can be appreciated around the nanodisks due to the conductive layer utilized to improve the SEM images, subsequently removed prior to dark-field spectroscopy measurements. The white scale bar represents 100

nm. (c) Schematic of the experimental setup. The incident white light transmitted through the sample without scattering (shown in cyan) is blocked in the intermediate back focal plane (BFP). The scattered light is shown in yellow. The dark-field image of a single nanocylinder is formed in the image plane (IP) at the entrance slit of the spectrometer.

**Realization of non-Huygens transparent metasurfaces**

Our novel effect can be harnessed to design fully transmissive, all-dielectric metasurfaces without relying on the well-known Huygens condition[66,67]. Contrarily to the latter, the light traverses the structure without significant phase variation, thus rendering the metasurface truly invisible (see **Figure 6**c-d). This is a direct consequence of Eq.(1). It can be easily seen by writing the transmission coefficient as a sum of the relevant multipole contributions of the meta-atoms[68]:

$$t = 1 + t_p + t_m + t_{Q^{(m)}} + t_{Q^{(e)}} \tag{6}$$

Each term in the previous sum is proportional to the corresponding total multipole moment (basic and toroidal contributions). For example, the contribution of the total electric quadrupole radiation to the transmission when the system is illuminated by x-polarized light is simply:

$$t_{Q^{(e)}} = \kappa \left[ Q_{xz}^{(e)} + \frac{ik_d}{v_d} T_{xz}^{(e)} \right] \tag{7}$$

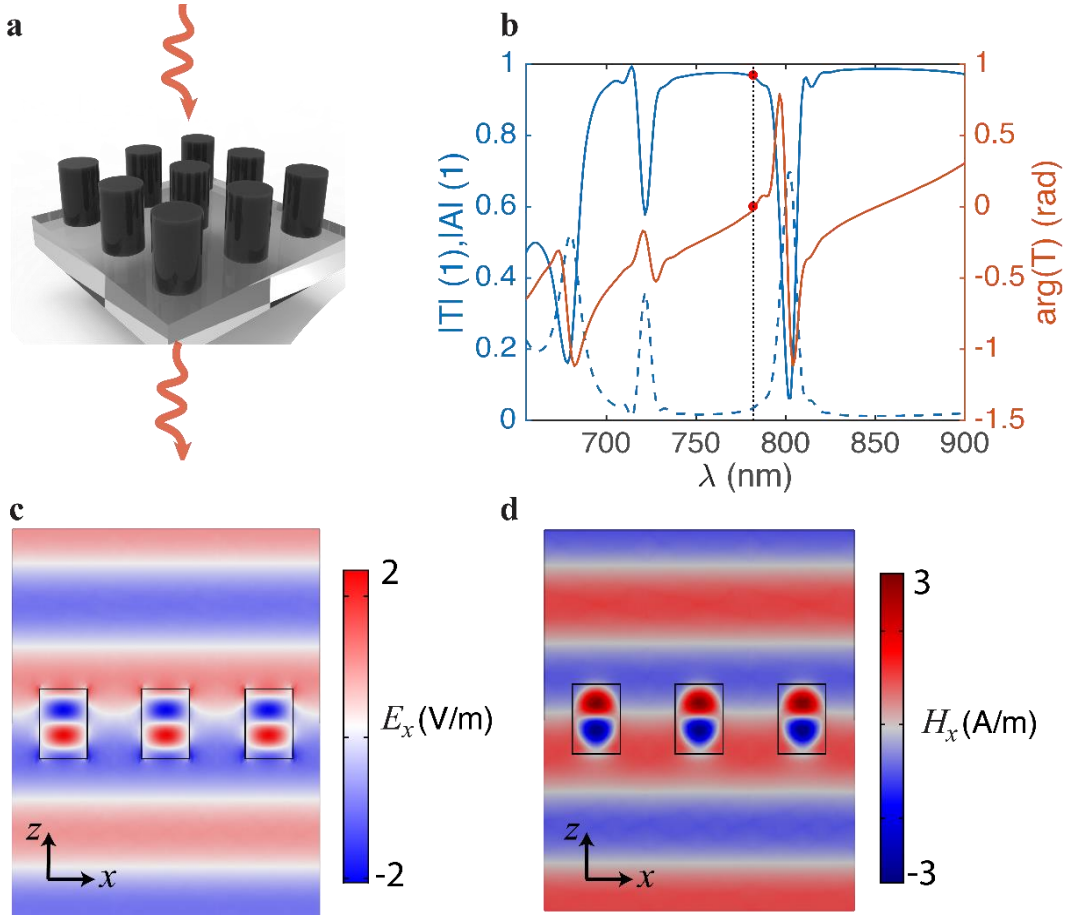

**Figure 6.** Design of a highly compact hybrid anapole-based metasurface. The spacing between the nanocylinders is set to 300 nm, i.e. at a value much smaller than the incident wavelength. (a) artistic representation of the proposed metasurface operating at the hybrid anapole state. The structure is illuminated by an x-polarized plane wave, which passes through the array completely unperturbed. (b)

Transmission, absorption and phase of the transmitted wave with respect to the incident one. As predicted by Eq. (1), a zero-phase difference at the hybrid anapole wavelength ($\lambda = 782$ nm), is observed (red dot). (c) x component of the electric field in the metasurface at the hybrid anapole state, when the incident plane wave amplitude is $E_0 = 1$ V/m. (d) y component of the magnetic field.

where $\kappa$ is a constant depending on the lattice period, incident wavelength and refractive index of the host environment[68]. When Eq. (1) is fulfilled, the term in brackets in Eq. (7) is exactly zero, and the same takes place for the other multipoles. Consequently, we are left with $t = 1$ in Eq. (6), and the incident wave leaves the system unperturbed (see **Figure 6**). Importantly, due to the multipolar suppression, the near field of each individual nanocylinder is exceptionally well confined (see e.g. **Figure 4**b). As a result, near field lattice coupling is strongly minimized, and the hybrid anapole state can be observed in very dense metasurfaces, such as the one schematized in **Figure 6**a, where the cylinder spacing is 0.4 times the incident wavelength. The considered period is not unique, i.e. once the geometry supporting the hybrid anapole for an isolated particle is known, a subwavelength metasurface of such particles will mimic the single particle behavior far away from the first diffraction order[69]. Noteworthy, the proposed metasurface allows a fast switch between total transparency (at the hybrid anapole) and an almost perfect absorption point (at the resonant wavelength of the $TE_{110}$ and $TM_{113}$ modes responsible for the hybrid anapole), which could be potentially useful for energy harvesting applications. Finally, we note that the results presented above are in remarkable contrast with the optical behavior of a solid aSi thin film at the hybrid anapole wavelength (reflection higher than 70%).

**Conclusion**

In summary, we have obtained a non-trivial invisibility state in a non-spherical all-dielectric particle with dimensions comparable to the incident wavelength. For the first time to our knowledge, we have been able to describe the present configuration as a combination of anapole states originating from different terms in the Cartesian charge-current multipole expansion of the particle, and thus demonstrated simultaneous cancellation of all leading contributions to scattering. Furthermore, the effect has been shown to be a consequence of the resonant interaction of a Fabry Perot and a Mie-like modes excited inside the scatterer, which are responsible for strong internal electric and magnetic hotspots, yielding an average of approximately 10 times energy density enhancement inside the particle, consequently –energy storing in the stationary regime. The hybrid anapole state is then shown both numerically and analytically to be "protected" against significant changes in the substrate refractive index. In this regard, we observe and explain a transition from a four-fold to a conventional electric dipole anapole as a consequence of the nature of the modes involved. While only two modes are responsible for an anapole state (see Supplementing Information), the hybrid anapole is originated by the interplay of several modes (resonant and background-like). This suggests a different scattering response in the transient regime, which could prove potentially useful for ultrafast computing and signal modulation.

We have confirmed our predictions by the first experimental measurement in the visible range of a dark hybrid anapole nanoparticle. Finally, we have demonstrated that the effect can be exploited for the realization of truly invisible (both in amplitude and phase) and dense metasurfaces in the visible range. The same resonant modes responsible for the hybrid anapole mediate strong absorption (almost 70%) in the spectral vicinity of the invisible state, a feature that could be potentially exploited in switchable energy harvesting devices.

The latter results are not only of fundamental interest but also suggest a route towards enhanced nonlinear effects in the absence of an elastic scattering background. We anticipate that these states can also be obtained in the visible range for non-centrosymmetric high-index materials such as AlGaAs or GaP, thus providing an ideal platform for both third and second harmonic sources.

More importantly, in this context the hybrid electric and magnetic nature of the modes at play could lead to new interesting interference phenomena between the high order nonlinear currents. Our results can be extended beyond nanophotonics and into the microwave realm, where hybrid anapoles, as 'ideal' nonradiating sources, could be exploited in future highly efficient wireless power transfer devices making use of near field coupling between a chain of subwavelength resonators supporting the effect.

Our theoretical and experimental findings broaden our knowledge of anapole electrodynamics beyond the electric dipole approximation, revealing new perspectives in the field and giving for the first time a solid physical picture in terms of the fundamental QNMs driving the resonator response. The results of this work will open new pathways for the design of efficient scattering suppression and sensing devices in the field of all-dielectric nanophotonics.